\documentclass[twoside]{article}
\usepackage{fleqn,espcrc2}
\usepackage{graphicx}
\title{Small-scale structure of cold dark matter}

\author{Dominik J. Schwarz\address{Institut f\"ur Theoretische Physik,
        Technische Universit\"at Wien, \\
        Wiedner Hauptstra\ss e 8--10, A-1040 Wien, Austria}
        \thanks{Supported by the Alexander von Humboldt foundation and the
        Austrian Academy of Sciences.}
        and 
        Stefan Hofmann\address{Institut f\"ur Theoretische Physik,
        Universit\"at Frankfurt, \\
        Postfach 11 19 32, D-60054 Frankfurt am Main, Germany}}
\begin{document}

\begin{abstract}
We investigate the clumping of cold dark matter (CDM) at small
scales. If the CDM particle is the neutralino, we find that collisional 
damping during its kinetic decoupling from the radiation fluid and 
free streaming introduce a small-scale cut-off in the primordial 
power spectrum of CDM. This cut-off sets the scale for the very first CDM 
objects in the Universe, which we expect to have a mass of $\sim 10^{-12} 
M_\odot$. For non-thermal CDM candidates, such as axions, wimpzillas, or 
primordial black holes, the cosmological QCD transition might induce features
in the primordial spectrum at similar mass scales.
\vspace{1pc}
\end{abstract}

\maketitle

\section{Introduction}

Dark matter with an equation of state that has been non-relativistic 
($p \ll \rho$) since structure formation started (around matter-radiation 
equality) is called cold dark matter \cite{CDM}. CDM gives rise to the 
hierarchical formation of large structures ($> 1$ Mpc) in the Universe, 
i.e.~small structures form first and grow to larger structures later.
The growth of CDM 
density fluctuations is suppressed during the radiation dominated epoch. Thus 
the rms CDM density fluctuations go like $k^3$ at large scales, but increase 
only logarithmically with wavenumber at scales well below the Hubble 
scale at matter-radiation equality. 

At very small scales ($\ll 1$ Mpc), the power spectrum and the evolution of 
CDM density fluctuations has not been discussed in detail so far, although 
the understanding of the small-scale behavior of CDM density fluctuations is 
essential for a realistic estimate of expected rates in CDM searches. The 
reason is that analytic calculations in this deeply non-linear regime are 
very challenging and numerical simulations do not have the dynamical range 
to resolve scales as small as the solar system. A priori one can say that 
there has to be a cut-off in the CDM power spectrum at some very small scale, 
otherwise the energy density in the fluctuations itself would be infinity. 

The nature of CDM is unknown. Here we mention three popular candidates with 
very different properties: The first one is the lightest neutralino,
which probably is the 
lightest supersymmetric particle \cite{LSP}. Its mass is 
expected to be in the range $40$ GeV  -- $600$ GeV \cite{Ellis}. The neutralino
is a mixture of the neutral gauginos and higgsinos, thus it interacts 
through weak interactions only. Below we assume that the lightest neutralino 
is the bino, because in the constrained minimal supersymmetric standard model 
the dominant contribution in the mix comes from the bino \cite{Ellis}. 

A second CDM candidate is the axion \cite{axion}. The axion mass has been 
restricted to the be $10^{-6}$ eV -- $10^{-2}$ eV \cite{Hagmann,axion} and 
contributes to the dark matter if the mass is small. Axions interact 
much weaker than weakly interacting particles, the interactions are 
suppressed by the Peccei-Quinn scale, which is $\sim 10^{12}$ GeV. 
Thus axions are never in thermal equilibrium with the radiation fluid. 
An example of small-scale structure in CDM has been found from the initial 
misalignment mechanism of axions by Hogan and Rees \cite{Hogan}. It turns out 
that, if the Peccei-Quinn scale is below the reheating temperature after 
inflation, large isocurvature perturbations in the axion density are created 
once the axion mass is switched on during the QCD transition. It has 
been shown that axion mini-clusters with $\sim 10^{-12} M_\odot$ and radii 
of $\sim 0.1 R_\odot$ might emerge, which may be observed by means of pico- 
and femtolensing \cite{femtolensing}. 

Let us mention a third CDM candidate: primordial black holes \cite{PBH}. Their 
mass should be $>10^{-16} M_\odot$ in order to survive until today \cite{Carr}.
Primordial black holes interact with the rest of the Universe via gravity only.
They may be found or excluded by gravitational lensing 
\cite{microlensing,femtolensing}. 

\section{Damping scales for CDM}

For any thermal CDM species there are two mechanisms that contribute to 
the damping of small-scale fluctuations: During the kinetic decoupling of CDM 
from the radiation fluid the mean free path is finite and thus collisional 
damping occurs. After the kinetic decoupling has been completed free streaming
can further wash out the remaining fluctuations. 

For neutralinos, chemical freeze out happens at $\sim m/20 > 2$ GeV
\cite{LSP}. In 
contrast kinetic decoupling happens at much smaller temperatures \cite{SSW2}, 
because elastic interactions with the radiation fluid are 
possible at temperatures as low as $1$ MeV. However, due to the large momentum 
of the neutralinos, many collisions are needed for a significant change of the 
momentum. It turns out that $N \sim m/T$ collisions can keep the neutralinos in
kinetic equilibrium and thus the relaxation time can be estimated as $\tau \sim
N \tau_{\rm coll}$, where the collision time for a bino is given by 
\begin{equation}
\tau_{\rm coll} \approx \left[5.5 \left( G_{\rm F} M_W^2\over M^2 
- m^2\right)^2 \; T^5\right]^{-1}\ , 
\end{equation}
with $M$ being the slepton mass and $m$ the mass of the bino. With a
slepton mass $M = 200$ GeV and a bino mass of $m = 100$ GeV, the 
relaxation time is given by $\tau \sim (10 \mbox{\ MeV}/T)^4 t_{\rm H}$, 
where $t_{\rm H}$ is the Hubble time. Kinetic decoupling of neutralinos 
happens at $T \sim 10$ MeV. 

We incorporate dissipative phenomena by describing the CDM as an imperfect 
fluid \cite{Weinberg}. The coefficients
of heat conduction, shear and bulk viscosity are estimated to be $m \chi \sim 
\eta \sim \zeta \sim n T \tau$, where $n$ is the number density of CDM 
particles. We find that the damping of density perturbations goes as 
\begin{equation}
\delta \propto \exp\left[ - \left(M_{\rm D}\over M\right)^{0.3}\right] \ ,
\end{equation} 
where the damping scale depends on the mass of the neutralino and the
slepton masses and is typically $M_{\rm D} = 10^{-13} M_\odot$ -- 
$10^{-10} M_\odot$. For comparison, the CDM mass within a Hubble patch is
$\sim 10^{-4} M_\odot$ at $T \sim 10$ MeV. 

Free streaming leads to additional damping. The velocity of neutralinos
right after kinetic decoupling is $v \sim (T/m)^{1/2} \sim 10^{-2}$.  
Free streaming also gives rise to exponential damping, due to the velocity
dispersion. The typical free steaming scale is $10^{-12} M_\odot$  -- 
$10^{-10} M_\odot$. Thus both damping 
mechanisms operate approximately at the same scale. The power spectrum of 
neutralino CDM is cut off at $M < 10^{-12} M_\odot$.

The mechanism of collisional damping also works for CDM in the form of
a heavy neutrino ($\sim 1$ TeV), but it does not work for wimpzillas 
\cite{wimpzilla}, because these are too heavy to ever be in thermal 
equilibrium. Free streaming induces a cut-off in the power spectrum for 
all mentioned CDM candidates. The scale and the strength of the damping 
depend on the masses and the primordial velocity distributions.

\section{QCD induced CDM clumps}

Besides damping mechanisms there are also processes that might enhance 
the primordial CDM spectrum at small scales. One is the formation of 
axion mini-clusters \cite{Hogan} that we mentioned already in the 
introduction. 

Together with Schmid and Widerin one of the present authors has found that 
large amplifications of density fluctuations might be induced by the 
QCD transition at scales $10^{-20} M_\odot < M < 10^{-10} M_\odot$ 
\cite{SSW1,SSW2}. The mechanism is the following: During a first-order 
QCD transition the sound speed vanishes. Thus the density perturbations 
of the dominant radiation fluid go into free fall and create large peaks 
and dips in the spectrum, which grow at most linearly with the wavenumber.
These peaks and dips produce huge gravitational potentials. CDM falls into 
these gravitational wells. It is important to note that this amplification 
mechanism works for matter that is kinetically decoupled at the QCD 
transition around $150$ MeV. For neutralinos this is not the case. Large 
inhomogeneities in the neutralinos would be washed out by collisional 
damping later on. A structure similar to the acoustic peaks in the 
photon-baryon fluid might survive. The large inhomogeneities in the radiation 
fluid are completely washed out during neutrino decoupling at $\sim 1$ MeV.

\section{The first CDM objects}

The smallest scales that survive damping are the first scales that go 
non-linear, thus these scales form the first gravitationally bound objects 
(apart from primordial black holes, if they exist) in the Universe. Let us 
estimate their size if the CDM is the neutralino. The scale is given by the 
cut-off at $M \sim 10^{-12} M_\odot$. With a COBE normalized \cite{COBE}
CDM spectrum we find the rms density fluctuations at equality to be
\begin{equation}
{\delta \rho\over \rho} \sim 2 \times 10^{-4} 
\left[\ln\left(k_{\rm D}\over k_{\rm eq}\right)\right]^{3/2} 
\left(k_{\rm D}\over k_{\rm eq}\right)^{(n-1)/2} \ ,
\end{equation}
which is $2 \times 10^{-2} (10^{-1})$ for the spectral index $n= 1(1.2)$. 
Thus these objects go nonlinear at or shortly after equality, at a redshift 
of $\sim 10^2 (10^3)$. If we assume that these clouds are spherical they 
would collapse to a radius of $10^3 (10^2) R_\odot$ today, which is by chance 
a very interesting scale for observations. Although $10^{-12} M_\odot$
in a volume of $(10^2 R_\odot)^3$ seems to be an extremely diluted cloud,
the overdensity of such a cloud today would be $\sim 10^{11}$.
In more optimistic scenarios (larger tilt and/or additional peaks in the 
spectrum from the QCD transition) it is even possible that these
clouds are so compact that pico- and femtolensing can be used to search 
for them.

It is unclear whether some of these first objects can survive up to 
today, this will be the subject of further studies. We think that 
understanding and revealing the small-scale structure of CDM might help us
to learn more about the nature of CDM.

\end{document}